\documentclass[]{aa}

\usepackage{natbib}
\bibpunct{(}{)}{;}{a}{}{,} % to follow the A&A style
\usepackage{graphicx}
\usepackage{txfonts}
\begin{document}

   \title{Clouds in the atmospheres of extrasolar planets}
   \subtitle{IV. On the scattering greenhouse effect of CO$_2$ ice particles:\\
             Numerical radiative transfer studies}

   \author{D. Kitzmann
          \inst{1},
          A.B.C. Patzer
          \inst{1},
          H. Rauer
          \inst{1,2}
          }
          
   \authorrunning{D. Kitzmann et al.}
   \titlerunning{Clouds in the atmospheres of extrasolar planets. IV.}

   %\offprints{D. Kitzmann}

   \institute{Zentrum f\"ur Astronomie und Astrophysik, Technische Universit\"at Berlin,
              Hardenbergstr. 36, 10623 Berlin (Germany)\\
              \email{kitzmann@astro.physik.tu-berlin.de}
              \and
              Institut f\"ur Planetenforschung, Deutsches Zentrum f\"ur Luft- und Raumfahrt (DLR),
              Rutherfordstr. 2, 12489 Berlin (Germany)
             }

   \date{Received 16 July 2012 / Accepted 17 June 2013}

% \abstract{}{}{}{}{} 
% 5 {} token are mandatory
 
  \abstract
   {Owing to their wavelengths dependent absorption and scattering properties, clouds have a strong impact on the climate of planetary atmospheres. Especially, the potential greenhouse effect of CO$_2$ ice clouds in the atmospheres of terrestrial extrasolar planets is of particular interest because it might influence the position and thus the extension of the outer boundary of the classic habitable zone around main sequence stars. Such a greenhouse effect, however, is a complicated function of the CO$_2$ ice particles optical properties.}
   {We study the radiative effects of CO$_2$ ice particles obtained by different numerical treatments to solve the radiative transfer equation. To determine the effectiveness of the scattering greenhouse effect caused by CO$_2$ ice clouds the radiative transfer calculations are performed over the relevant wide range of particle sizes and optical depths employing different numerical methods.}
   {We use Mie theory to calculate the optical properties of particle polydispersion. The radiative transfer calculations are done with a high order discrete ordinate method (DISORT). Two-stream radiative transfer methods are used for comparison with previous studies.}
   {The comparison between the results of a high-order discrete ordinate method and simpler two-stream approaches reveals large deviations in terms of a potential scattering efficiency of the greenhouse effect. The two-stream methods overestimate the transmitted and reflected radiation, thereby yielding a higher scattering greenhouse effect. For the particular case of a cool M-type dwarf the CO$_2$ ice particles show no strong effective scattering greenhouse effect by using the high-order discrete ordinate method, whereas a positive net greenhouse effect was found in case of the two-stream radiative transfer schemes. As a result, previous studies on the effects of CO$_2$ ice clouds using two-stream approximations overrated the atmospheric warming caused by the scattering greenhouse effect. Consequently, the scattering greenhouse effect of CO$_2$ ice particles seems to be less effective than previously estimated. In general, higher order radiative transfer methods are necessary to describe the effects of CO$_2$ ice clouds accurately as indicated by our numerical radiative transfer studies.}
   {}


   \maketitle
%
%________________________________________________________________

\section{Introduction}

Clouds can have an important impact on the climate of terrestrial planetary atmospheres by either scattering the incident stellar radiation back to space (albedo effect) or by trapping the infrared
radiation in the atmosphere (greenhouse effect). The extension of the habitable zone around different types of stars depends, therefore, on the presence of clouds (see e.g. \citet{Kasting1993,Selsis2007}).
The position of the inner boundary of the habitable zone is determined by the efficiency of the albedo effect by water droplet clouds \citep{Kasting1988Icar}. The outer boundary, on the other hand, might be influenced by the formation of CO$_2$ ice clouds and their corresponding climatic impact \citep[e.g.][]{Kasting1993, Forget2010ASPC}. \citet{Selsis2007} discussed the potential effects of CO$_2$ clouds for the outer boundary of the habitable zone around the star Gliese 581. In particular the planet Gliese 581d was studied with one and three-dimensional atmospheric models by \citet{Wordsworth2010A&A,Wordsworth2011ApJ} which included a simplified model for the description of CO$_2$ clouds. Their results showed that the CO$_2$ clouds contribute to the greenhouse effect by increasing the surface temperature of Gliese 581d.

\begin{figure*}
  \centering
  \resizebox{\hsize}{!}{\includegraphics{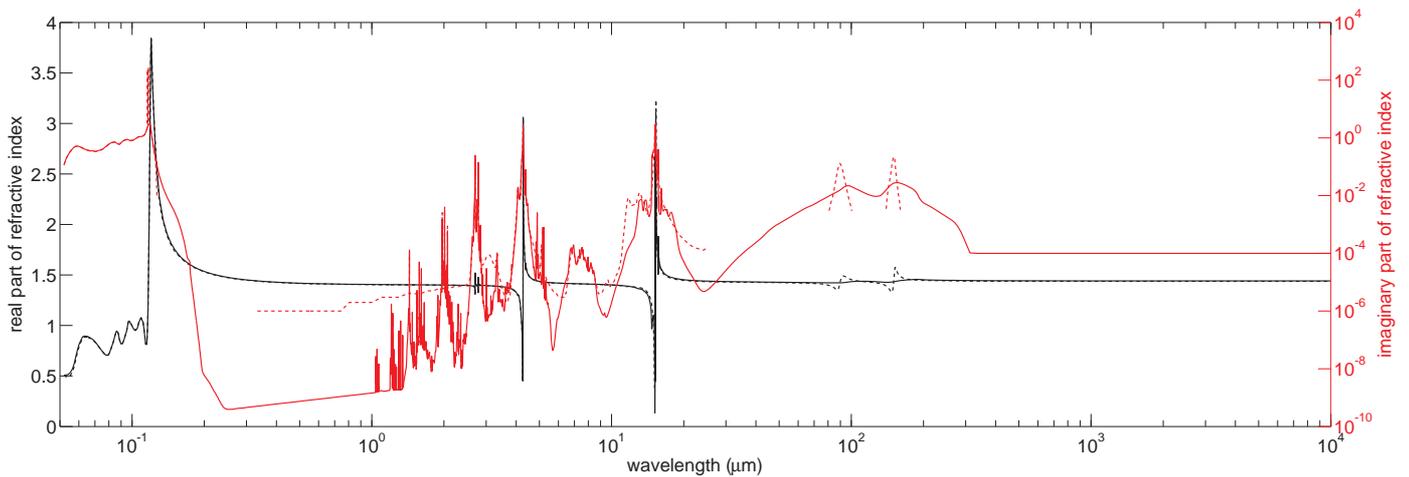}}
  \caption{Real part (black line) and imaginary part (red line) of the refractive index of CO$_2$ ice. The solid lines indicate the data published by \citet{Hansen1997JGR,Hansen2005JGRE}, the dashed lines mark the data compilation from \citet{Warren1986ApOpt}.}
  \label{fig:co2data}
\end{figure*}

The effects of CO$_2$ ice clouds, however, differ from those of water droplet or ice clouds due to the very different optical properties. Additionally, whether the greenhouse or albedo effect dominates for a given cloud of CO$_2$ ice crystals depends on a variety of cloud properties. Apart from the wavelength-dependent optical depths, the crystal size is the most important quantity which determines the climatic effect of CO$_2$ clouds \citep{Pierrehumbert1998JAtS,Forget1997}.

The only CO$_2$ clouds observed so far in planetary atmospheres are found in the atmosphere of Mars. Most of the observed clouds near the Martian equator are composed of particles with rather small effective radii ($a_{\mathrm{eff}} < 3 \ \mathrm{\mu m}$) and optical depths smaller than 0.5 in the visible wavelength region \citep{Maattanen2010Icar,Montmessin2007JGRE,Vincendon2011JGRE}. Larger particles ($a_{\mathrm{eff}} \sim 10 \ \mathrm{\mu m}$) are found at high latitudes during the polar night \citep{Forget1995JGR,Colaprete2003JGRE}, for example.

Clouds composed of CO$_2$ ice crystals are also thought to play a major role for the conditions in the early Martian atmosphere. For a dense early Mars atmosphere \citet{Forget1997} have shown that CO$_2$ ice clouds can exhibit a greenhouse effect by scattering of infrared radiation if the effective particle radii are larger than $\sim6 \ \mathrm{\mu m}$. Smaller particle radii yield a dominating albedo effect which leads to lower surface temperatures. A similar approach for early Mars, but including a more detailed radiative transfer treatment, was used by \citet{Mischna2000Icar}. Their results indicate that the altitude of CO$_2$ clouds influence the efficiency of the greenhouse effect. The formation of CO$_2$ clouds in the atmosphere of early Mars has been studied by \citet{Colaprete2003} with a detailed microphysical model using laboratory data from \citet{Glandorf2002Icar} for the description of the CO$_2$ ice nucleation. The microphysical model was coupled with a time-dependent one-dimensional radiative-convective atmospheric model. According to their model calculations, the mean particle size in a dense CO$_2$ dominated atmosphere of early Mars is of the order of $1000 \ \mathrm{\mu m}$. The results of \citet{Colaprete2003} also indicate that the (scattering) greenhouse effect is overall limited by the evaporation of CO$_2$ ice particles due to the large amount of latent heat released during the their formation. The greenhouse effect of such CO$_2$ clouds is thereby self-limited. Besides early Mars, CO$_2$ clouds may have also contributed to the climate of the early Earth \citep[see e.g.][]{Caldeira1992Natur}.

In this paper we study the radiative effects of CO$_2$ ice particles. Mie theory is used to calculate the optical properties for assumed gamma particle size distributions for a broad range of effective radii (Sect. \ref{sec:mie theory}). In Sect. \ref{sec:radiative_transfer} discrete ordinate radiative transfer schemes are applied to obtain the spectral reflectance and transmittance of given CO$_2$ clouds for different optical depths. The net radiative effects of such CO$_2$ ice clouds for the particular case of a cool M-type dwarf star are discussed in Sect. \ref{sec:co2_effects}.
To make a comparison with previous model studies on the climatic effects of CO$_2$ ice clouds we also apply two-stream radiative transfer methods. These additional calculations are compared with our findings obtained from a high-order discrete ordinate method.

\section{Optical properties of CO$_\mathbf{2}$ ice particles}
\label{sec:mie theory}

\subsection{Refractive index of CO$_2$ ice}
\label{sec:refractive_indices}

Two different (main) compilations of the refractive index are available for CO$_2$ ice covering the wavelength range from the far UV up to the microwave region and are shown in Fig. \ref{fig:co2data}. The first set of this refractive index was published by \citet{Warren1986ApOpt}. The imaginary part was compiled from various published measurements by different authors, connected by extrapolation and interpolation. Gaps in the data are present in the VIS and FIR wavelength range (see Fig. \ref{fig:co2data}) where no published measurements were available at that time. Based on the compiled imaginary part, \citet{Warren1986ApOpt} calculated the corresponding real part by using the Kramers-Kronig relations \citep{Kramers1927,Kronig1926JOSA}.

The second compilation was published by \citet{Hansen1997JGR,Hansen2005JGRE}. From $0.174 \ \mathrm{\mu m}$ up to $333 \ \mathrm{\mu m}$ laboratory measurements by Hansen were used for the imaginary part of the refractive index. Below $0.174 \ \mathrm{\mu m}$ and above $333 \ \mathrm{\mu m}$ the compilation was extended with data from \citet{Warren1986ApOpt}. The corresponding real part of the refractive index was also obtained via the Kramers-Kronig relations.

The imaginary part of the refractive index is directly related to the absorption coefficient. For CO$_2$ it shows considerable variations of about twelve orders of magnitudes throughout the whole wavelength range. Especially in the IR it exhibits a complicated oscillatory behaviour with several strong absorption bands. Besides these strong bands the imaginary part suggests that the overall absorption of CO$_2$ ice seems to be small in the IR in comparison to e.g. water ice (see \citet{Warren08} for details on the refractive index of H$_2$O ice). Compared to the imaginary part, the real part does not show such large variations. It, however, has two distinct large features in the IR near $4.3 \ \mathrm{\mu m}$ and $15.2 \ \mathrm{\mu m}$.

Note, that besides these two large data compilations, other measurements are also available for the astrophysically important strong absorption bands in the IR and FIR. The refractive index for these bands was published by e.g. \citet{Johnson1996Icarus}, \citet{Hudgins1993ApJS}, \citet{Ehrenfreund1996A&A}, or \citet{Baratta1998JOSAA}. The wavelengths positions of these absorption bands agree overall with those of the \citet{Hansen1997JGR,Hansen2005JGRE} and \citet{Warren1986ApOpt} data compilations. However, the bands widths and heights are strong functions of temperature and can, therefore, vary between the different measurements.
For this study the refractive index from \citet{Hansen1997JGR,Hansen2005JGRE} is used.

\subsection{Mie theory calculations}
\label{sec:optical_properties}

To calculate the optical properties of CO$_2$ ice particles Mie theory \citep{Mie1908AnP} is used in this study assuming a spherical particle shape. From observations in the Earth atmosphere it is well known that water ice crystals are rarely spherical but show a broad distribution of different particle shapes. In contrast to H$_2$O the unit cell of CO$_2$ ice has a face-centred cubic structure. Therefore, as found e.g. in in-situ laboratory measurements by \citet{Behnken1912}, \citet{Wahl1913}, or \citet{Wergin1997} CO$_2$ crystals can have cubic or octahedral shapes. Combinations of both (cuboctahedra) or more complicated shapes such as rhobic-dodecahedral crystals also occur. In-situ measurements of the shapes of CO$_2$ ice cloud particles in, for example, the Martian atmosphere are not yet available. Furthermore, the scattering properties of such non-spherical particles are extremely demanding to calculate so that we use the approximation of spherical particles in this study which has also been done in all previous studies on the climatic impact of CO$_2$ ice clouds.

Following the approach of \citet{Forget1997} we describe the size distribution of the cloud particles by a modified gamma distribution:

\begin{equation}
  f(a) = \frac{\left( a_\mathrm{eff} \nu \right)^{2-1/\nu}}{\Gamma\left(\frac{1-2\nu}{\nu}\right)} a^{\left(\frac{1}{\nu}-3\right)} \mathrm e^{-\frac{a}{a_\mathrm{eff}\nu}} \,
  \label{eq:gamma_distribution}
\end{equation}
where $a_\mathrm{eff}$ is the effective radius, $\nu$ the effective variance, and $\Gamma$ the gamma function. 
For the effective variance we adopt the value $\nu = 0.1$ also used by \citet{Forget1997}. The effective radii are varied between $0.1 \ \mathrm{\mu m}$ and $200 \ \mathrm{\mu m}$  in this study. 
Mie theory \citep{Wiscombe1980ApOpt,Wiscombe1979msca} is used to obtain the optical properties for a distinct single particle size. These optical properties are then averaged over the assumed size distribution functions.

\begin{figure}
  \centering
  \resizebox{\hsize}{!}{\includegraphics{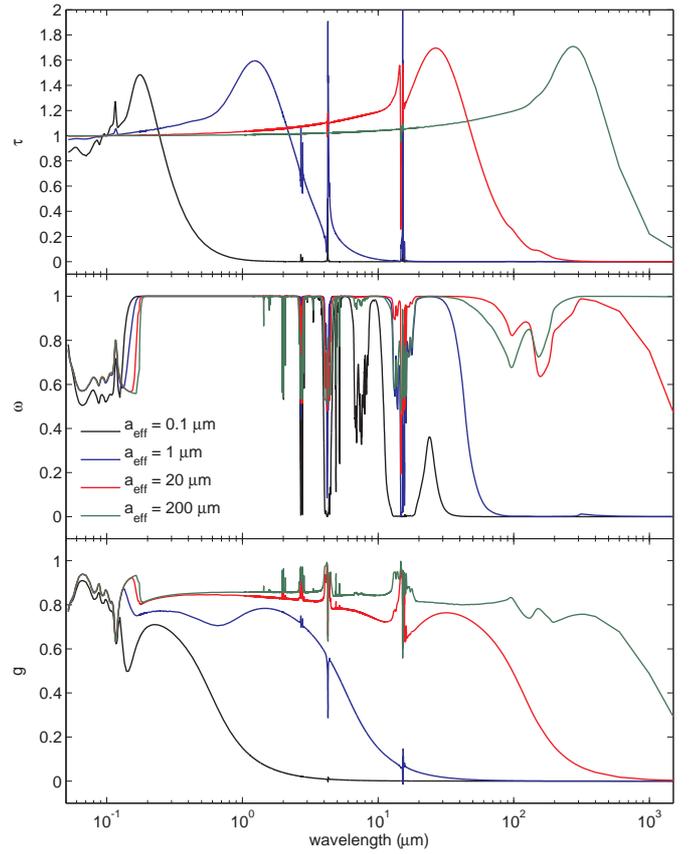}}
  \caption{Calculated optical properties of CO$_2$ for four different size distributions: $a_{\mathrm{eff}}=0.1 \ \mathrm{\mu m}$ (\textit{black line}), $a_{\mathrm{eff}}=1 \ \mathrm{\mu m}$ (\textit{blue line}), $a_{\mathrm{eff}}=20.0 \ \mathrm{\mu m}$ (\textit{red line}), $a_{\mathrm{eff}}=200 \ \mathrm{\mu m}$ (\textit{green line}). \textit{Upper diagram}: optical depth, \textit{middle diagram}: single scattering albedo $\omega$, \textit{lower diagram}: asymmetry parameter $g$.}
  \label{fig:optical_properties}
\end{figure}

The optical properties from Mie scattering calculations are shown in Fig. \ref{fig:optical_properties} for four different exemplary effective particle radii and an optical depth\footnote{Unless otherwise stated $\tau$ refers to the particular wavelength of $\lambda = 0.1 \ \mathrm{\mu m}$.} of $\tau=1$.

The optical depth of the larger particles is almost constant from the EUV up to the EHF wavelength region. This corresponds directly to the large particle limit of Mie theory \citep{BohrenHuffman1998asls} which predicts a constant value for the extinction efficiency independent of the refractive index and, therefore, from the considered material. In case of the largest particles ($a_{\mathrm{eff}} = 200\ \mathrm{\mu m}$) this limit is reached for almost the whole wavelength range up to the FIR (Fig. \ref{fig:optical_properties}). For each size distribution the optical depths show a distinct maximum at different wavelengths. This maximum value is located near the wavelengths roughly corresponding to the particle size. For example, the particles with $a_{\mathrm{eff}}=20 \ \mathrm{\mu m}$ ($a_{\mathrm{eff}}=1 \ \mathrm{\mu m}$) have their maximum optical depth in the IR at $\lambda=27 \ \mathrm{\mu m}$ ($\lambda=1.2 \ \mathrm{\mu m}$). The positions of these maxima will largely determine the radiative effects of the CO$_2$ ice particles.

The single scattering albedo is shown in the middle panel of Fig. \ref{fig:optical_properties}. The resulting albedo values indicate that large CO$_2$ ice crystals are scattering dominant over a wide wavelengths range. Independent of the particle size, the single scattering albedo is almost unity from the near UV up to the NIR. As a direct consequence the incident stellar radiation will mostly be scattered and not absorbed at this wavelength range.

The albedo of CO$_2$ ice in the IR shows a complicated behaviour as a function of the effective particle size. Within the strong absorption bands the albedo can be as low as zero, indicating dominating absorption. However, outside of these bands the scattering albedo of the $200\ \mathrm{\mu m}$ and $20\ \mathrm{\mu m}$ particle size distributions are almost unity up to the EHF and FIR, respectively. The particles with $a_{\mathrm{eff}}=0.1 \ \mathrm{\mu m}$ show much stronger variations in the single scattering albedo than the distributions with larger effective radii. In the IR, however, the single scattering albedo quickly approaches a value of almost zero for these small particles. 

The single scattering albedo emphasises the fact that the interaction of the radiation field with CO$_2$ ice particles is quite different compared to e.g. water ice clouds. While H$_2$O ice clouds exhibit a greenhouse effect by absorption and re-emission of thermal radiation, CO$_2$ ice particles will affect the thermal radiation mostly by scattering processes. In order to produce a corresponding (scattering) greenhouse effect \citep{Forget1997} the thermal radiation must be scattered back to the planetary surface by the CO$_2$ ice clouds. Thus, the efficiency of the greenhouse effect of CO$_2$ ice depends also on the mean scattering angle and, therefore, on the value of the asymmetry parameter.

\begin{figure*}
  \centering
  \resizebox{\hsize}{!}{\includegraphics{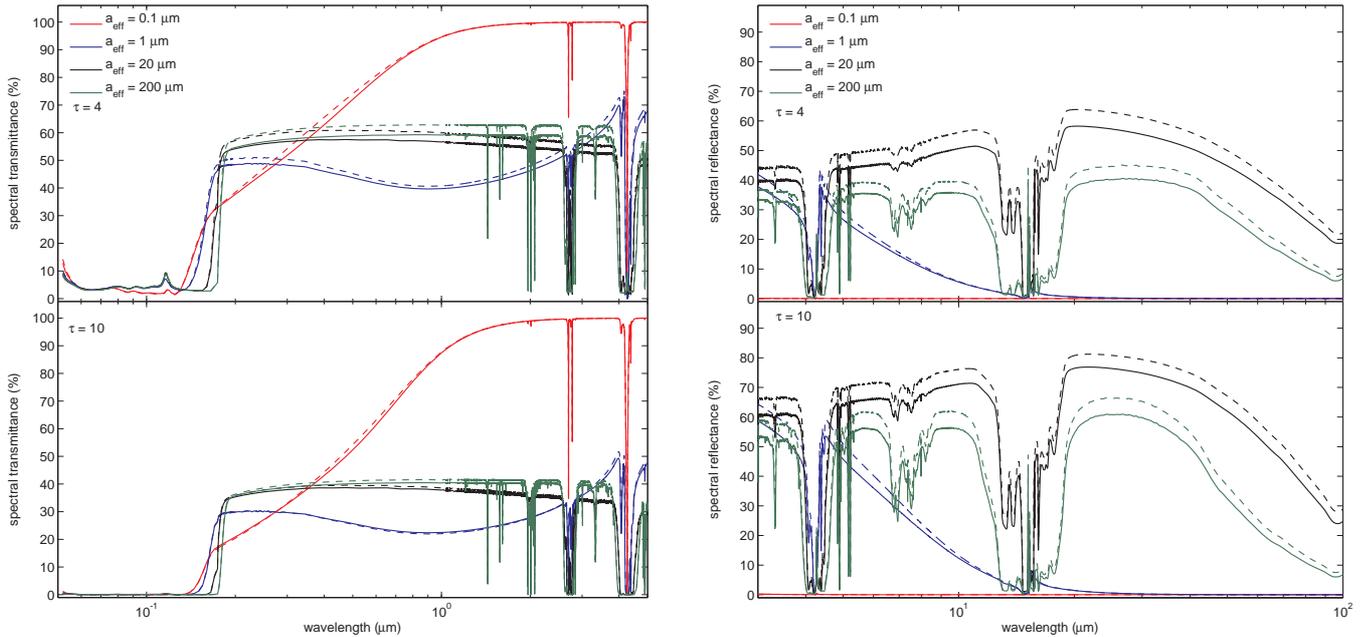}}
  \caption{Spectral transmittance (\textit{left column}) and reflectance (\textit{right column}) for two different optical depths: $\tau=4$ (\textit{upper panels}) and $\tau = 10$ (\textit{lower panels}). Results are shown for four different size distributions: $a_{\mathrm{eff}}=0.1 \ \mathrm{\mu m}$ (\textit{red line}), $a_{\mathrm{eff}}=1 \ \mathrm{\mu m}$ (\textit{blue line}), $a_{\mathrm{eff}}=20.0 \ \mathrm{\mu m}$ (\textit{black line}), and $a_{\mathrm{eff}}=200 \ \mathrm{\mu m}$ (\textit{green line}). Calculations using DISORT (two-stream methods) are denoted by solid lines (dashed lines).}
  \label{fig:transmittance_reflectance}
\end{figure*}

The asymmetry parameters averaged over the size distributions are shown in the lower panel of Fig. \ref{fig:optical_properties}. The resulting asymmetry parameters are always larger than or equal to zero (Rayleigh scattering limit), which indicates dominating forward scattering. For the largest particles with $a_{\mathrm {eff}} = 200 \ \mathrm{\mu m}$ the asymmetry parameter has almost a constant value of about $0.9$ from the UV up to the FIR except for strong variations in the absorption bands. With a value of about $0.8$ the asymmetry for $a_{\mathrm {eff}} = 20 \ \mathrm{\mu m}$ is smaller in the wavelength regime from the UV up to the NIR. It then steadily increases up to $0.9$ in the MIR before slowly declining for wavelength $\lambda > 30 \ \mathrm{\mu m}$. Only for wavelengths larger than $300 \ \mathrm{\mu m}$ the asymmetry parameter approaches the limit of Rayleigh scattering ($g = 0$) as expected. The smallest particles considered here ($a_{\mathrm {eff}} = 0.1 \ \mathrm{\mu m}$) show a constant decline in $g$ already at $\lambda = 0.3 \ \mathrm{\mu m}$ and approach the Rayleigh limit in the NIR.

\section{Radiative transfer calculations}
\label{sec:radiative_transfer}

For each particle size distribution radiative transfer calculations are performed to determine the radiative effects of the considered CO$_2$ ice particles. In these calculations only one single cloud layer is studied, assuming plane-parallel geometry. 

We consider the radiative transfer equation

\begin{equation}
   \mu \frac{\mathrm{d} I_\lambda}{\mathrm{d}\tau_\lambda} = I_\lambda - S_{\lambda,\mathrm{*}}(\tau_\lambda) 
    - \omega_\lambda \frac{1}{2} \int_{-1}^{+1} p_\lambda(\mu,\mu') I_\lambda(\mu') \mathrm{d}\mu'
	  \label{eq:rte}
\end{equation}
with the scattering phase function $p_\lambda$, the single scattering albedo $\omega_\lambda$, and the contribution due to an external illumination by a central star $S_{\lambda,\mathrm{*}}(\tau_\lambda)$. The scattering phase function can be represented as an infinite series of Legendre polynomials \citep{Chandrasekhar1960ratr}
\begin{equation}
  p_\lambda(\mu,\mu') = \sum_{n=0}^{\infty} (2 n + 1) P_n(\mu) P_n(\mu') \chi_{\lambda,n}
\end{equation}
with the Legendre polynomials $P_n(\mu)$ and the phase function moments $\chi_{\lambda,n}$. In practice the series is truncated at a certain $n=N_\mathrm{max}$. For discrete ordinate methods the number of moments $N_\mathrm{max}$ used to describe the phase function expansion series is a direct function of the number of ordinates (streams) considered \citep{Chandrasekhar1960ratr}.

In this study the scattering phase function is approximated by the Henyey-Greenstein function \citep{Henyey1941ApJ....93...70H}. The Henyey-Greenstein phase function depends only on the asymmetry parameter $g$ and is an approximation to the full Mie scattering phase function. Although lacking many detailed features and complicated structure of the Mie phase function, the Henyey-Greenstein function preserves its average quantities, most notably its asymmetry parameter. The Henyey-Greenstein phase function is usually a good replacement for the much more complicated Mie phase function in case of higher optical depths because the finer details of the Mie phase functions are effaced due to the multiple scattering effects. Only the averaged quantities (such as the asymmetry parameter) then determine the overall properties of the radiation field, especially the angular integrated quantities of the intensity, such as the radiation flux. Note, however, that at very small optical depths the results obtained by using the Henyey-Greenstein function may not be so accurate.

For the solution of the radiative transfer equation Eq. (\ref{eq:rte}) we employ different discrete ordinate methods \citep{Chandrasekhar1960ratr}. In particular, we use the well established and very flexible atmospheric radiative transfer code DISORT \citep{Stamnes1988ApOpt}. Following the previous studies of \citet{Mischna2000Icar} or \citet{Colaprete2003} we additionally also use two-stream methods for comparison, namely a $\delta$-Eddington quadrature method for the incident stellar radiation and a hemispheric two-stream method in the IR (see \citet{Toon1989JGR....9416287T} for details on these methods).

The simple two-stream solution methods use only one stream for the upward and one for the downward direction, respectively. In this case only the first two moments of the phase function expansion would enter into the radiative transfer calculations. However, even the most simple phase functions, such as the Rayleigh scattering phase function, require the first three moments of the expansion for an accurate representation. Therefore, the two-stream methods such as the $\delta$-Eddington quadrature method or the hemispheric mean two-stream radiative transfer, cannot use the usual expansion series but must rely on different, simplified approximations of the phase function. For example, the hemispheric mean method uses a phase function of $1+g$ in the forward and $1-g$ in the backward direction \citep{Toon1989JGR....9416287T}.

For the application of the more general discrete ordinate solver DISORT we use here 24 streams in all calculations, i.e. $N_\mathrm{max} + 1 = 24$. This allows for a much better description of the phase function than in the case of the two-stream methods. Such an approach usually should give more accurate results for systems dominated by angular dependent radiative transfer processes such as in particular anisotropic scattering.

\begin{figure}
  \centering
  \resizebox{\hsize}{!}{\includegraphics{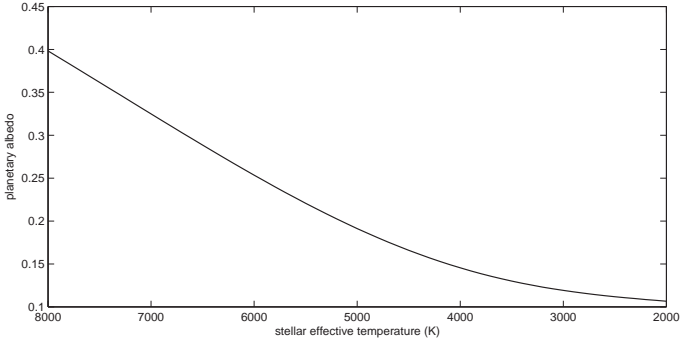}}
  \caption{Planetary albedo (clear-sky condition) as a function of the effective temperatures of the central stars.}
  \label{fig:planetary_albedo}
\end{figure}

In order to make an estimate, the scattering and absorption behaviour of atmospheric CO$_2$ molecules and of the planetary surface has been approximately included in the radiative transfer calculations for the stellar radiation with a wavelength-dependent albedo $a_{\mathrm{p},\lambda}$ below the cloud layer. The albedo $a_{\mathrm{p},\lambda}$ roughly describes the impact of an Earth-like surface with a 2 bar CO$_2$ atmosphere as an example. Therefore, we assume a constant (measured) Earth-like surface albedo of 0.13 \citep{Kitzmann2010A&A...511A..66K}, lowered to 0.1 to account for the NIR absorption of CO$_2$ molecules. In addition to this constant albedo describing the impact of the planetary surface we added a Rayleigh-scattering like albedo with the well-known $\lambda^{-4}$ dependence to account for molecular Rayleigh scattering. The Rayleigh scattering contribution is important for central stars with higher effective temperatures whereas the influence of the (reduced) surface albedo is important for cooler stars. Figure \ref{fig:planetary_albedo} shows the resulting (clear-sky) planetary albedo $a_\mathrm{p}$ as a function of the effective temperatures of the central stars. 
We note that the surface albedo, and the amount of gas which influences the Bond albedo due to Rayleigh scattering can have a large impact on the net effect of a CO$_2$ cloud. For example, low surface albedos (e.g. due to oceans) will yield a smaller net greenhouse effect while a more reflective, Martian-like surface albedo can result in an increase of the scattering greenhouse effect. As mentioned by \citet{Forget1997} the thermal emission by the cloud itself would have only a minor contribution and is therefore neglected in our study.

\begin{figure}
  \centering
  \resizebox{\hsize}{!}{\includegraphics{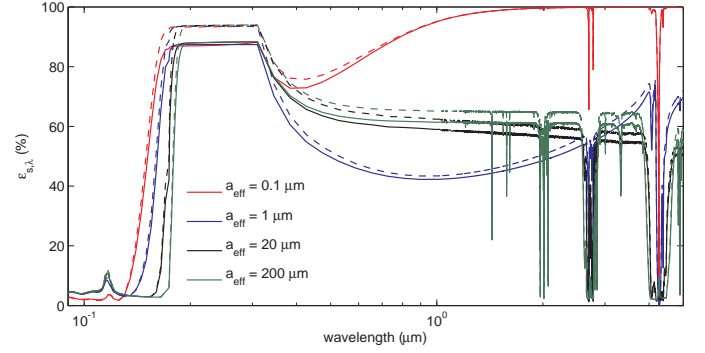}}
  \caption{Calculated fraction of the incident stellar radiation in downward direction at the cloud base ($\epsilon_\mathrm{s,\lambda}$) for $\tau = 4$. Results are shown for four different size distributions: $a_{\mathrm{eff}}=0.1 \ \mathrm{\mu m}$ (\textit{red line}), $a_{\mathrm{eff}}=1 \ \mathrm{\mu m}$ (\textit{blue line}), $a_{\mathrm{eff}}=20.0 \ \mathrm{\mu m}$ (\textit{black line}), and $a_{\mathrm{eff}}=200 \ \mathrm{\mu m}$ (\textit{green line}). Calculations using DISORT (two-stream methods) are denoted by solid lines (dashed lines).}
  \label{fig:downward_stellar_radiation}
\end{figure}

A zenith angle for the incident stellar radiation of 60 degrees is used here which corresponds to the mean zenith angle in a one-dimensional global average atmospheric model. As a result of our radiative transfer calculations we obtain the fraction of the incident stellar radiation in downward direction $\epsilon_\mathrm{s,\lambda}$ at the cloud base for each wavelength $\lambda$ by
\begin{equation}
  \label{eq:ratio_star}
  \epsilon_\mathrm{s,\lambda} = \frac{F_{\mathrm{s,cb},\lambda}^\downarrow}{F_{\mathrm{s},\lambda}^\downarrow}
\end{equation}
where $F_{\mathrm{s},\lambda}^\downarrow$ denotes the incident stellar radiation at the top of the atmosphere and $F_{\mathrm{s,cb},\lambda}^\downarrow$ the downward shortwave radiation flux at the cloud base. The flux $F_{\mathrm{s,cb},\lambda}^\downarrow$ is composed of the incident stellar light which is transmitted and forward scattered through the cloud layer, as well as contributions due to multiple scattering of that shortwave radiation between the cloud base and the lower atmosphere/planetary surface (approximated here by the wavelength-dependent albedo $a_{\mathrm{p},\lambda}$). Additionally, we also obtain the spectral transmittance of the cloud layer and the percentage of thermal IR radiation scattered back towards the surface (spectral reflectance $\epsilon_\mathrm{a,r,\lambda}$ of the cloud) given by
\begin{equation}
  \label{eq:ration_ir}
  \epsilon_\mathrm{a,r,\lambda} = \frac{F_{\mathrm{a,cb},\lambda}^\downarrow}{F_{\mathrm{a},\lambda}^\uparrow}
\end{equation}
where $F_{\mathrm{a},\lambda}^\uparrow$ is the thermal radiation flux emitted from the atmosphere/surface and $F_{\mathrm{a,cb},\lambda}^\downarrow$ denotes the downward longwave flux at the cloud base. Note that the incident fluxes for each wavelength are set to unity to simplify the numerical boundary conditions in the radiative transfer calculations. Obviously, this does not affect the computed values of the ratios $\epsilon_\mathrm{a,r,\lambda}$ and $\epsilon_\mathrm{s,\lambda}$.

To illustrate the basic differences between the 2-stream methods and DISORT Fig. \ref{fig:transmittance_reflectance} summarises the resulting spectral transmittance and reflectance of the CO$_2$ cloud for several different cloud properties. Additionally, the ratio $\epsilon_\mathrm{s,\lambda}$ is depicted in Fig. \ref{fig:downward_stellar_radiation}.

The smallest particles ($a_{\mathrm {eff}} = 0.1 \ \mathrm{\mu m}$) yield an almost 100\% transmittance for wavelengths larger than $1 \ \mathrm{\mu m}$ even for a high optical depth of 10. This implies that their effect on the incident stellar radiation will be small if the maximum of the stellar radiation is located in this wavelength region. On the other hand, they also have a reflectance of almost zero which means that these small particles won't yield a scattering greenhouse effect (see Fig. \ref{fig:transmittance_reflectance}).

The transmittance of larger particles is always less than 100\% and decreases at higher optical depth. The strongest negative impact on the transmittance is obtained for particles with $a_{\mathrm {eff}} = 1 \ \mathrm{\mu m}$ because these particles have their largest contribution to the opacity in this wavelength region (see Fig. \ref{fig:optical_properties}). The transmittance is almost equal for the largest particles. This is the result of the large particle limit discussed in Sect. \ref{sec:optical_properties} yielding the same optical properties (optical depth, asymmetry parameter, and single scattering albedo) for all big particles at short wavelengths.

The spectral reflectance for the $a_{\mathrm {eff}} = 1 \ \mathrm{\mu m}$ particles is very low (Fig. \ref{fig:transmittance_reflectance}), especially in the wavelength region where most of the atmospheric thermal radiation is transported ($\lambda > 8 \ \mathrm{\mu m}$). This and their large negative impact on the transmittance suggests that these particles will rather cool the lower atmosphere than cause a net scattering greenhouse effect.

However, larger particles can show high reflectance values in the IR wavelength range. The reflectance for $a_{\mathrm {eff}} = 20 \ \mathrm{\mu m}$ is overall almost 20\% higher than for the largest particles with $a_{\mathrm {eff}} = 200 \ \mathrm{\mu m}$. This is caused by the higher optical depth of the $20 \ \mathrm{\mu m}$ particles in the thermal IR wavelength range (cf. Fig. \ref{fig:optical_properties}) and, on the other hand, by the higher asymmetry parameter for $a_{\mathrm {eff}} = 200 \ \mathrm{\mu m}$ which leads to a larger fraction of the thermal radiation being scattered in forward direction to space rather than backwards to the planetary surface. This indicates that the scattering greenhouse effect will be most efficient for particles with sizes comparable to the wavelength of the thermal radiation.

\begin{figure*}
  \centering
  \resizebox{0.6\hsize}{!}{\includegraphics{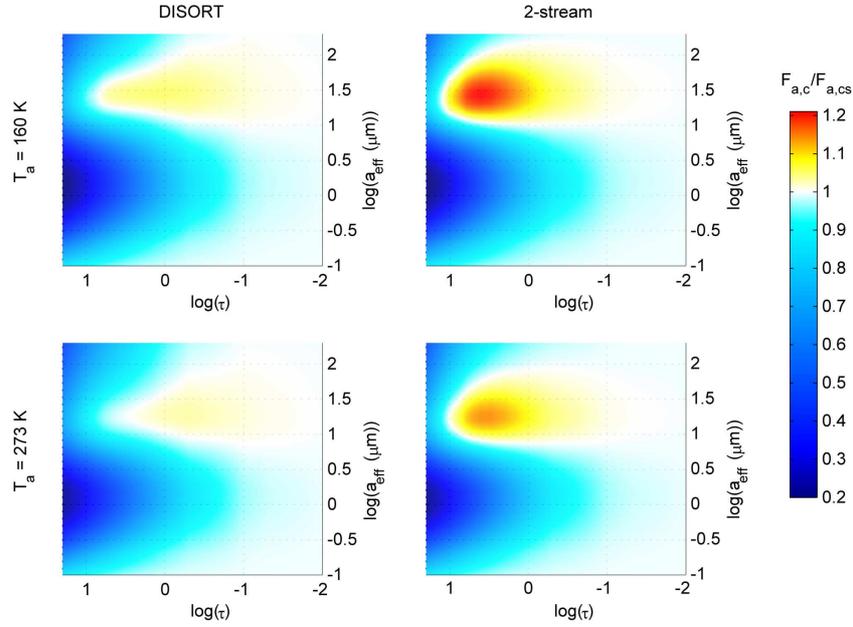}}
  \caption{Ratios $F_\mathrm{a,c}/F_\mathrm{a,cs}$ as a function of optical depth $\tau$ and the effective radius $a_\mathrm{eff}$ of the gamma size distribution for an effective temperature of $T_\mathrm{eff} = 3000 \ \mathrm{K}$ and two values of the atmospheric temperatures $T_a$: 160 K (\textit{upper diagrams}), 273 K (\textit{lower diagrams}). Results are shown for calculations using DISORT (\textit{left column}) and two-stream methods (\textit{right column}). Notice, that the same colour scaling is used for each diagram.}
  \label{fig:forcing_mstar}
\end{figure*}

Figs. \ref{fig:transmittance_reflectance} and \ref{fig:downward_stellar_radiation} also show the results of the two-stream methods. The results indicate that the two-stream methods overestimate both:
\begin{itemize}
  \item the transmission/scattering of stellar radiation 
  \item and the back-scattering of thermal radiation. 
\end{itemize}	
For the $\delta$-Eddington two-stream method the differences are about 6\% while the hemispheric mean two-stream yields deviations up to 20\%. Thus, two-stream methods will in general overate the positive net scattering greenhouse effect of CO$_2$ clouds by allowing more stellar radiation to be transmitted through the cloud layer and more thermal radiation back-scattered to the planetary surface. However, the exact climatic impact will depend on the spectral distributions of the radiation incident on the cloud. These effects are therefore discussed for different central stars and atmospheric temperatures in the following

\section{Radiative impact of CO$_\mathbf{2}$ clouds}
\label{sec:co2_effects}

By using the results of the radiative transfer calculations with normalised boundary conditions presented in the previous section we now determine the net radiative effect of a given CO$_2$ cloud for different incident stellar spectra and atmospheric temperatures.

Let $F_{\mathrm{s},\lambda}^\downarrow$ denote the downward flux from the central star incident at the top of the atmosphere which can be written as
\begin{equation}
  F_{\mathrm{s},\lambda}^\downarrow = f_{\mathrm{s},\lambda} F_{\mathrm{s}}^\downarrow
  \label{eq:flux_splitup}
\end{equation}
where $f_{\mathrm{s},\lambda}$ is the normalised spectral distribution and $F_{\mathrm{s}}^\downarrow$ is the total (wavelength-integrated) flux. For simplicity we describe $F_{\mathrm{s},\lambda}^\downarrow = F_{\mathrm{s},\lambda}^\downarrow(T_{\mathrm{eff}})$ by black-body radiation with a given stellar effective temperature\footnote{Note, that non-black-body spectra could, of course, be studied straightforwardly.} $T_{\mathrm{eff}}$. Likewise, the upward thermal radiation incident at the cloud base is denoted by $F_{\mathrm{a},\lambda}^\uparrow$ which can also be factorised in analogy to Eq. (\ref{eq:flux_splitup}):
\begin{equation}
  F_{\mathrm{a},\lambda}^\uparrow = f_{\mathrm{a},\lambda} F_{\mathrm{a}}^\uparrow \ ,
  \label{eq:flux__ir_splitup}
\end{equation}
where $F_\mathrm{a}^\uparrow$ is the wavelength-integrated flux and $f_{\mathrm{a},\lambda}$ the corresponding normalised spectral distribution. The infrared radiation from the lower atmosphere $F_{\mathrm{a},\lambda}$ is assumed to be black-body radiation of a given temperature $T_\mathrm{a}$. This temperature $T_\mathrm{a}$ is not necessarily the surface temperature of the planet but could be any temperature below the cloud base where the atmosphere becomes transparent in the IR.

In the cloud-free case the total amount of incident shortwave radiation being absorbed by the planet is determined as usual by $(1-a_\mathrm{p}) F_{\mathrm{s}}^\downarrow$, where $a_\mathrm{p}$ is the (cloud-free) planetary albedo (see Fig. \ref{fig:planetary_albedo}) given by
\begin{equation}
  a_\mathrm{p}(T_{\mathrm{eff}}) = \int_0^\infty a_{\mathrm{p},\lambda} f_{\mathrm{s},\lambda} \mathrm d \lambda \ .
\end{equation} 
Conservation of energy implies that this absorbed energy is balanced by the emitted (clear sky - cs) thermal flux $F_{\mathrm{a,cs}}^\uparrow$:
\begin{equation}
  \label{eq:balance_cs}
  F_\mathrm{a,cs}^\uparrow = (1 - a_\mathrm{p}) F_\mathrm{s}^\downarrow \ .
\end{equation}

In contrast to a clear sky, the amount of shortwave radiation absorbed by the planet in the cloudy case (c) is given by:
\begin{eqnarray}
  \int_0^\infty ( 1 - a_\mathrm{p,\lambda}) F_{\mathrm{s,cb},\lambda}^\downarrow \mathrm d \lambda 
  & = & \int_0^\infty ( 1 - a_\mathrm{p,\lambda}) \epsilon_\mathrm{s,\lambda} F_{\mathrm{s},\lambda}^\downarrow \mathrm d \lambda \nonumber \\
  & = & F_{\mathrm{s}}^\downarrow \int_0^\infty ( 1 - a_\mathrm{p,\lambda}) \epsilon_\mathrm{s,\lambda} f_{\mathrm{s},\lambda} \mathrm d \lambda \ ,
\end{eqnarray}
using Eq. (\ref{eq:ratio_star}) for $\epsilon_\mathrm{s,\lambda}$ and Eq. (\ref{eq:flux_splitup}) for $F_{\mathrm{s},\lambda}^\downarrow$.

Thus, for each considered stellar effective temperature we calculate the total percentage $\epsilon_\mathrm{s}$ of the incident shortwave radiation being absorbed by the planet according to:
\begin{equation}
  \epsilon_\mathrm{s}(T_{\mathrm{eff}}) = \int_0^\infty (1 - a_{\mathrm{p},\lambda}) \epsilon_\mathrm{s,\lambda} f_{\mathrm{s},\lambda} \mathrm d \lambda \ ,
\end{equation}
using an adaptive Gauss-Kronrod quadrature method.

On the other hand, the amount of thermal radiation scattered back towards the surface by the cloud is given by (see Eqs. (\ref{eq:ration_ir}) \& (\ref{eq:flux__ir_splitup}))
\begin{equation}
  \int_0^\infty F_{\mathrm{a,cb},\lambda}^\downarrow \mathrm d \lambda 
   = \int_0^\infty \epsilon_\mathrm{a,r,\lambda} F_{\mathrm{a,c},\lambda}^\uparrow \mathrm d \lambda 
   = F_{\mathrm{a,c}}^\uparrow \int_0^\infty \epsilon_\mathrm{a,r,\lambda} f_{\mathrm{a},\lambda} \mathrm d \lambda \ .
\end{equation}

Therefore, the total percentage $\epsilon_{a,r}$ of the upwelling thermal flux $F_{\mathrm{a,c}}^\uparrow$ back-scattered by the cloud is obtained by
\begin{equation}
  \epsilon_\mathrm{a,r}(T_\mathrm{a}) = \int_0^\infty \epsilon_\mathrm{a,r,\lambda} f_{\mathrm{a},\lambda} \mathrm d \lambda \ .
\end{equation}
Note, that the wavelength region ($14 \ \mathrm{\mu m} - 16 \ \mathrm{\mu m}$) around the $15 \ \mathrm{\mu m}$ absorption band of CO$_2$ is excluded here in the integration. Both, the CO$_2$ gas molecules and the CO$_2$ cloud can strongly absorb and re-emit thermal radiation in that particular spectral region which might result in a classical greenhouse effect. Because the cloud's thermal emission as well as the absorption and thermal emission by CO$_2$ molecules are not considered in our radiative transfer calculations (see previous section), this wavelength region is omitted from the analysis of the scattering greenhouse effect. Calculations by e.g. \citet{Forget1997} which included also the CO$_2$ gas, however, showed that the contribution of this spectral region to the greenhouse effect of the CO$_2$ cloud seems to be small.

According to the conservation of energy, the upwelling thermal flux in the cloudy case $F_{\mathrm{a,c}}^\uparrow$ is balanced by the absorbed downward shortwave flux $F_{\mathrm{s,cb}}^\downarrow$ and back-scattered thermal flux $F_{\mathrm{a,cb}}^\downarrow$:
\begin{eqnarray}
  F_\mathrm{a,c}^\uparrow &=& \int_0^\infty (1- a_\mathrm{p,\lambda}) F_{\mathrm{s,cb,\lambda}}^\downarrow \mathrm d \lambda 
                            + \int_0^\infty F_{\mathrm{a,cb},\lambda}^\downarrow \mathrm d \lambda \\
                           &=& \epsilon_\mathrm{s} F_\mathrm{s}^\downarrow + \epsilon_\mathrm{a,r} F_\mathrm{a,c}^\uparrow \nonumber
                          = (1 - \epsilon_\mathrm{a,r})^{-1} \epsilon_\mathrm{s} F_\mathrm{s}^\downarrow \label{eq:balance_cloudy} \ .
\end{eqnarray}

Therefore, we define the ratio of the thermal radiation fluxes of the cloudy ($F_\mathrm{a,c}$, Eq. (\ref{eq:balance_cloudy})) and clear-sky case ($F_\mathrm{a,cs}$, Eq. (\ref{eq:balance_cs})):
\begin{equation}
  \frac{F_\mathrm{a,c}}{F_\mathrm{a,cs}} = \frac{\epsilon_\mathrm{s}}{(1 - \epsilon_\mathrm{a,r})(1 - a_\mathrm{p})} \quad \left\{ 
    \begin{array}{ll} 
      > 1 \quad \mathrm{net \ \ heating \ \ effect}\\ 
      = 1 \quad \mathrm{radiatively \ \ neutral}\\ 
      < 1 \quad \mathrm{net \ \ cooling \ \ effect}
    \end{array} 
  \right.
\label{eq:cloud_forcing}
\end{equation}
which gives an indication of the net radiative effect of the cloud layer. For a ratio larger than one the cloud has a net heating effect on the atmosphere below the cloud, while for values smaller than one the lower atmosphere is cooled by the cloud layer. A ratio of exactly one represents radiatively neutral cloud particles.

To study the effectiveness of the (net) scattering greenhouse effect several calculations are performed for a broad range of stellar effective temperatures between 8000 K and 2000 K. Since cool M-type dwarf stars seem to be of particular importance in view of the detectability of (potentially) habitable planets \citep{Rauer2011A&A...529A...8R}, detailed results for a low effective stellar temperature of 3000 K are presented in the next subsection. 

For the thermal radiation we assume two limiting cases: first we consider the atmosphere below the cloud to be optically thin, such that $T_\mathrm{a}$ would correspond to the surface temperature of a habitable planet at the outer boundary of the habitable zone (i.e. $T_\mathrm{a} = 273 \ \mathrm{K}$, freezing point of water). In the second case we consider the atmosphere below the cloud to be opaque such that only thermal radiation directly from below the cloud base reaches the cloud. In this case we adopt a temperature $T_\mathrm{a} = 160 \ \mathrm{K}$ which roughly corresponds to the temperature where CO$_2$ would condense (cf. model calculations of \citet{Mischna2000Icar} or \citet{Colaprete2003}).

\subsection{Results for a cool M-dwarf star}

\begin{figure}
  \centering
  \resizebox{\hsize}{!}{\includegraphics{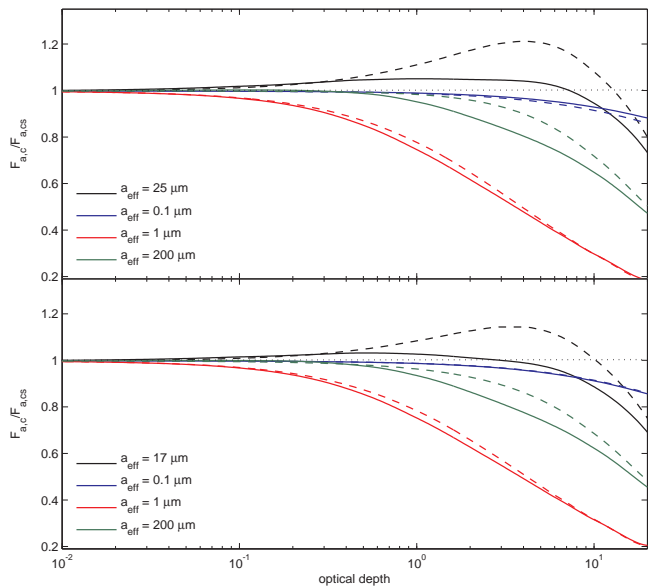}}
  \caption{Ratios $F_\mathrm{a,c}/F_\mathrm{a,cs}$ as a function of optical depth $\tau$ for an effective temperature of $T_\mathrm{eff} = 3000 \ \mathrm{K}$ and the two different atmospheric temperatures: $T_a = 160 \ \mathrm{K}$ (\textit{upper diagram}) and $T_a = 273 \ \mathrm{K}$ (\textit{lower diagram}). Results are shown for different size distributions in each diagram: $a_{\mathrm{eff}}=25 \ \mathrm{\mu m}$ (\textit{black line, upper diagram}), $a_{\mathrm{eff}}=17 \ \mathrm{\mu m}$ (\textit{black line, lower diagram}), $a_{\mathrm{eff}}=0.1 \ \mathrm{\mu m}$ (\textit{red line}), $a_{\mathrm{eff}}=1.0 \ \mathrm{\mu m}$ (\textit{blue line}), and $a_{\mathrm{eff}}=200 \ \mathrm{\mu m}$ (\textit{green line}). Calculations using DISORT (two-stream methods) are denoted by solid lines (dashed lines).}
  \label{fig:forcing_mstar_tau}
\end{figure}
\begin{figure}
  \centering
  \resizebox{\hsize}{!}{\includegraphics{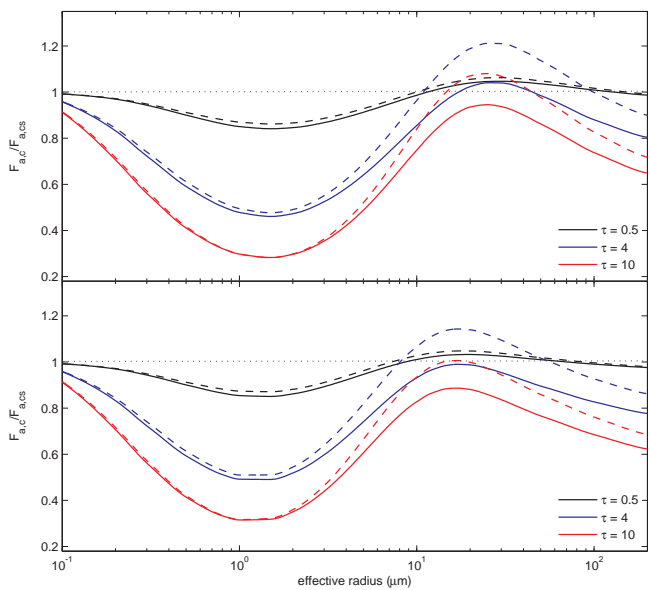}}
  \caption{Ratios $F_\mathrm{a,c}/F_\mathrm{a,cs}$ as a function of effective particle radius $a_{\mathrm{eff}}$ for an effective temperature of $T_\mathrm{eff} = 3000 \ \mathrm{K}$ and the two different atmospheric temperatures: $T_a = 160 \ \mathrm{K}$ (\textit{upper diagram}) and $T_a = 273 \ \mathrm{K}$ (\textit{lower diagram}). Results are shown for different values of the optical depth $\tau$ in each diagram: $\tau = 0.5$ (\textit{black line}), $\tau = 4$ (\textit{red line}), and $\tau=10$ (\textit{blue line}). Calculations using DISORT (two-stream methods) are denoted by solid lines (dashed lines).}
  \label{fig:forcing_mstar_size}
\end{figure}	
	
In the following we discuss our results for a stellar effective temperature of 3000 K, roughly corresponding to a cool M5 dwarf star. The ratios $F_\mathrm{a,c}/F_\mathrm{a,cs}$ are shown in Fig. \ref{fig:forcing_mstar} as a function of the optical depth $\tau$ and effective particle radius $a_\mathrm{eff}$ for both considered atmospheric temperatures. The results of two-stream radiative transfer calculations are also shown for comparison. For a more detailed analysis, different slices through the two-dimensional parameter space are shown in the Figs. \ref{fig:forcing_mstar_tau} and \ref{fig:forcing_mstar_size} for several different particle sizes and optical depths.

The results depicted in Figs. \ref{fig:forcing_mstar}, \ref{fig:forcing_mstar_tau}, and \ref{fig:forcing_mstar_size} suggest that the CO$_2$ particles have a negative or neutral impact over a large range of the considered parameter space. Only particles with sizes comparable to the wavelength of the thermal radiation contribute to a net greenhouse effect if the optical depth is not too large. For higher optical depths, again a cooling effect is found. This kind of behaviour as a function of the optical depth is consistent with the results of \citet{Mischna2000Icar} and \citet{Colaprete2003}.

Thus, CO$_2$ ice clouds have a certain parameter rage of particle radii and optical depths where the heating effect is most efficient. These particle radii are a function of the temperature $T_a$ and increase with decreasing temperature, i.e. the most efficient particle radius is about $25 \ \mathrm{\mu m}$ ($17 \ \mathrm{\mu m}$) for $T_a = 160 \ \mathrm{K}$ ($T_a = 273 \ \mathrm{K}$).
Apart from the shift in the particle size, also the efficiency increases with lower atmospheric temperature which leads to a stronger scattering greenhouse effect if the cloud layer would e.g. be located at higher altitudes within the atmosphere \citep[cf.][]{Mischna2000Icar}. However, even the highest ratios are only slightly larger than 1, i.e. the resulting quantitative radiative forcing would be small.

For larger or smaller particle sizes a cooling effect is found which is very large for particles with effective radii near $1 \ \mathrm{\mu m}$. This is caused by the large optical depth of these particles at small wavelengths where the maximum of the incident stellar radiation is located (see Sect. \ref{sec:radiative_transfer} and Fig. \ref{fig:transmittance_reflectance}). These kind of particles would lead to massive cooling if present in the atmosphere.

\begin{figure*}
  \centering
  \resizebox{0.6\hsize}{!}{\includegraphics{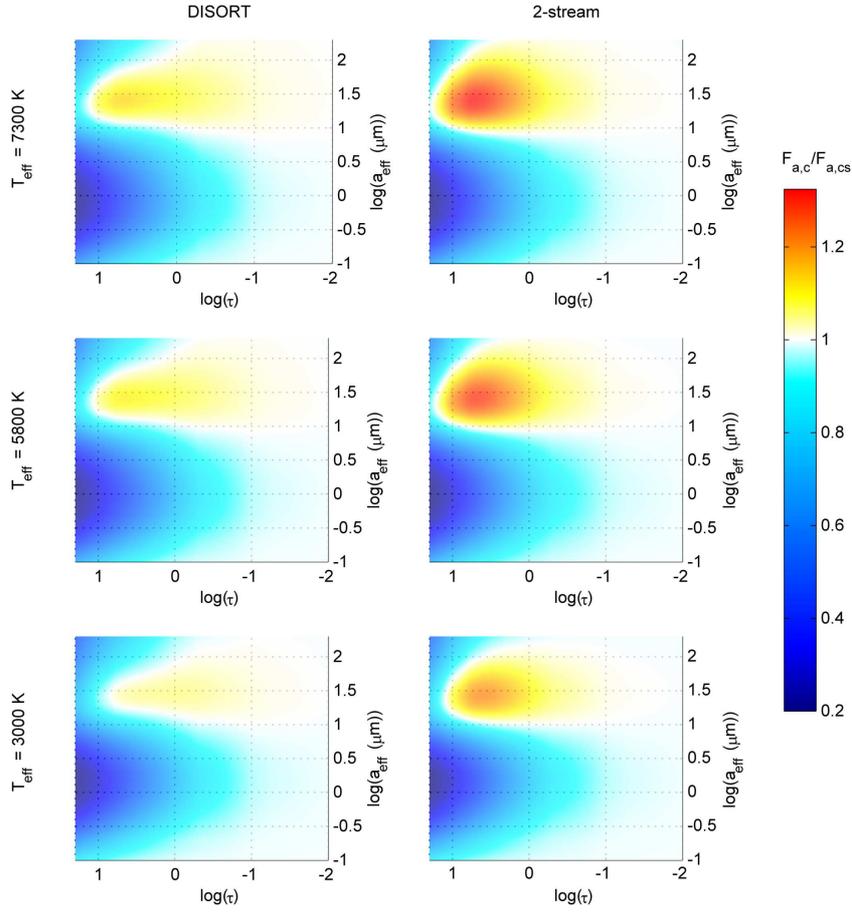}}
  \caption{Ratios $F_\mathrm{a,c}/F_\mathrm{a,cs}$ as a function of optical depth $\tau$ and the effective radius $a_\mathrm{eff}$ of the gamma size distribution for three different effective temperatures of the central star: 7300 K (\textit{upper diagrams}), 5800 K (\textit{middle diagrams}), and 3000 K (\textit{lower diagrams}). Results are shown for calculations using DISORT (\textit{left column}) and two-stream methods (\textit{right column}). Notice, that the same colour scheme is used in each diagram.}
  \label{fig:forcing_all}
\end{figure*}

Very small particles ($0.1 \ \mathrm{\mu m}$) are almost radiatively neutral, independent from the considered optical depth. Even at $\tau = 20$ they show only a small cooling effect. The reason for this behaviour is that these particles have their main contribution to the optical depth at very small wavelengths (cf. Sect. \ref{sec:radiative_transfer}). Since the maximum of the incident stellar radiation for the chosen low effective temperature is located near $\lambda = 1 \ \mathrm{\mu m}$ the impact of these small particles is very small. This effect is similar to the known low efficiency of Rayleigh scattering by the gas molecules for cool M-type stars.

According to the comparison of the different radiative transfer methods (Figs. \ref{fig:forcing_mstar}, \ref{fig:forcing_mstar_tau}, and \ref{fig:forcing_mstar_size}) the efficiency of warming by CO$_2$ ice clouds differs quite noticeably for different approximations in the numerical treatment solving the radiative transfer equation. Consistent with previous reports of e.g. \citet{Forget1997} or \citet{Mischna2000Icar} the two-stream methods predict a strong scattering greenhouse effect at medium optical depths near $\tau = 4$. In contrast to this, the application of DISORT, however, results in an almost radiatively neutral cloud. As already mentioned, only at small optical depths near 0.5 a slight positive effect is found which is much smaller than the two-stream results. Obviously the largest deviations occur for particle sizes which roughly correspond to the wavelength of the thermal radiation because this is the region where Mie scattering is important. The hemispheric mean two-stream method in particular seems to yield quite inaccurate results compared to a more elaborate discrete ordinate radiative transfer. The errors are also clearly a function of the optical depth. For lower optical depths the differences are small, whereas they are the largest in the region of the most efficient greenhouse effect. For more increasing optical depths the deviations decrease again.

\subsection{Other main sequence stars}
\label{sec:phase_functions}

\begin{figure}
  \centering
  \resizebox{\hsize}{!}{\includegraphics{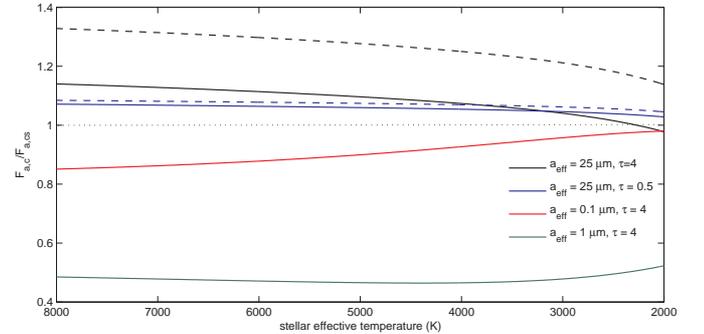}}
  \caption{Ratios $F_\mathrm{a,c}/F_\mathrm{a,cs}$ as a function of the stellar effective temperature. Results are shown for different values of the optical depth $\tau$ and effective radii of the gamma size distributions. \textit{Black line}: $a_{\mathrm{eff}}=25 \ \mathrm{\mu m}, \tau = 4$, \textit{blue line}: $a_{\mathrm{eff}}=25 \ \mathrm{\mu m}, \tau = 0.5$, \textit{red line}: $a_{\mathrm{eff}}=0.1 \ \mathrm{\mu m}, \tau = 4$, \textit{green line}: $a_{\mathrm{eff}}=1 \ \mathrm{\mu m}, \tau = 4$. Calculations using DISORT are denoted by solid lines, two-stream methods are marked with dashed lines.}
  \label{fig:forcing_2d_ms}
\end{figure}

Fig. \ref{fig:forcing_all} shows the ratio $F_\mathrm{a,c}/F_\mathrm{a,cs}$ for three different stellar effective temperatures: 7300 K (comparable to an F2V star), 5800 K (G2V star), and 3000 K (M5V star). All figures use the same colour scheme to make them directly comparable. Results are only shown for $T_a = 160 \ \mathrm{K}$. As noted in the previous subsection the values of $F_\mathrm{a,c}/F_\mathrm{a,cs}$ would be lower for higher atmospheric temperatures.
Additionally, Fig. \ref{fig:forcing_2d_ms} shows the resulting ratios as a function of the effective temperature for several chosen values of the particles effective radii and optical depths.

The results in Fig. \ref{fig:forcing_all} indicate that the most efficient particle size for a net scattering greenhouse effect does not vary with the type of the central star. Rather, it is only a function of the lower atmospheric temperatures $T_a$. As discussed for $T_a = 160 \ \mathrm{K}$ the corresponding effective particle size is about $25 \ \mathrm{\mu m}$. As shown in Fig. \ref{fig:forcing_2d_ms} this particle size yields a small net greenhouse effect for cool stars which increases towards higher effective temperatures. This increase is caused by the higher (clear-sky) planetary albedos due to the stronger Rayleigh scattering at shorter wavelengths.

Other particle sizes, however, exhibit larger variations as a function of the central star temperature. For example, lower ratios of $F_\mathrm{a,c}/F_\mathrm{a,cs}$ are obtained for increasing stellar effective temperatures in case of the particles with $a_{\mathrm {eff}} = 0.1 \ \mathrm{\mu m}$. For cooler stars they are almost radiatively neutral, whereas they show a strong cooling effect for higher stellar temperatures. This is directly related to the optical properties (see Fig. \ref{fig:optical_properties}) and the resulting transmittance from the radiative transfer calculations. As depicted in Fig. \ref{fig:transmittance_reflectance} a clear decline at shorter wavelength ($\lambda < 1 \ \mathrm{\mu m}$) is present in the spectral transmittance which makes the albedo effect more efficient at higher effective temperatures of the central star. The largest negative impact is still obtained for particle size distributions with effective radii close to $1 \ \mathrm{\mu m}$. This result is almost independent from the stellar effective temperatures.

As already discussed for the cool M-type star, the two-stream methods again strongly overestimate the positive effect of the CO$_2$ cloud. The ratios $F_\mathrm{a,c}/F_\mathrm{a,cs}$ obtained by DISORT are in every case smaller than those found by the two-stream methods. Especially at medium optical depths of $\tau = 4$ large deviations from the more accurate radiative transfer calculations are found. Here, the two-stream methods yield a large scattering greenhouse effect which also increases at higher stellar effective temperatures. For cooler stars, a less efficient greenhouse effect is obtained (see also \citet{Wordsworth2011ApJ}). At smaller optical depths, the differences between the two different radiative transfer methods are again smaller. These results indicate that the scattering greenhouse effect was most likely overrated in all previous model studies on the climatic impact of CO$_2$ clouds.

\section{Summary}

In this study we investigated the radiative effects of CO$_2$ ice clouds. The particle size distribution were assumed to be gamma distributions with different effective radii. Mie theory was used to calculate the optical properties (optical depths, single scattering albedo, and asymmetry parameter) of CO$_2$ ice particles with effective radii from $0.1 \ \mathrm{\mu m}$ to $200 \ \mathrm{\mu m}$. Radiative transfer calculations using different discrete ordinate methods were performed to obtain the spectral reflectance and transmittance of a single layer of CO$_2$ ice particles. In particular, a high order discrete ordinate solver (DISORT) was employed. Additionally, we also applied more simpler two-stream radiative transfer methods for comparison with previous studies. We then studied the net radiative effect of the CO$_2$ ice cloud for different spectral distributions of the incident stellar and the atmospheric thermal radiation.

A net scattering greenhouse effect by CO$_2$ clouds was only obtained for very specific cloud properties. Only particles with sizes comparable to the wavelength of the thermal radiation incident on the cloud base yielded a net greenhouse effect. For larger or smaller particles a net cooling effect was found. In particular, particles with $a_{\mathrm {eff}} \sim 1 \ \mathrm{\mu m}$ lead to strong cooling. Very small particles ($a_{\mathrm {eff}} = 0.1 \ \mathrm{\mu m}$) were found to be radiatively neutral for very cool stars even at high optical depths. In general, the net heating effect of the CO$_2$ cloud increases with increasing stellar effective temperatures. For cooler stars, however, even the most efficient particle sizes yield only a radiatively neutral cloud.

As suggested by Eq. (\ref{eq:cloud_forcing}) the planetary surface albedo and the albedo due to molecular Rayleigh scattering can have a large impact on the net climatic effect of a CO$_2$ cloud. While planets with low surface albedos (e.g. ocean planets) would result in a smaller net heating effect, planets with e.g. a Martian-like surface albedo can show an increased scattering greenhouse effect. Increasing the amount of CO$_2$ gas on the other hand would yield a stronger Rayleigh scattering but also more NIR absorption of the stellar radiation. Thus, for central stars with higher effective temperatures one can expect an increased net heating effect by the cloud while for cooler M-type stars the CO$_2$ cloud would remain more or less radiatively neutral.

We compared these results also with those of two-stream radiative transfer calculations. Here, a strong greenhouse effect was found at optical depth larger than 1 and smaller than 10 for particle sizes comparable to the wavelength of the thermal radiation which is in agreement with e.g. \citet{Forget1997} or \citet{Mischna2000Icar}. Overall, the two-stream methods yielded large deviations from the calculations using a higher order discrete ordinate method in the important parameter range where a net greenhouse effect was obtained in previous model studies.

Therefore, all these previous studies on the effects of CO$_2$ ice clouds on the outer boundary of the habitable zone which were restricted to two-stream approximations overestimated the positive scattering greenhouse effect. It is evident that more accurate radiative transfer methods are necessary to accurately describe the radiative effects of CO$_2$ clouds. Our study suggests that the scattering greenhouse effect still occurs. However, it is much smaller than previously expected and only efficient for a small range of parameters (optical depths, particle sizes). Therefore, we suggest to use higher order radiative transfer schemes in future atmospheric model calculations which include CO$_2$ ice clouds.

In combination with the reduced classical greenhouse effect by CO$_2$ gas molecules due to a revised description of the collisional induced absorption reported by \citet{Wordsworth2010Icar} this indicates that the outer boundary of the habitable zone should be located closer to the central star than previously considered (e.g. \citet{Selsis2007}). A quantitative analysis of the scattering greenhouse effect of CO$_2$ clouds and their impact on the position of the outer boundary of the habitable zone would, however, require an atmospheric model for more detailed calculations.

Note, that we used a zenith angle of 60 degree in this study which corresponds to the global average zenith angle for a one-dimensional model. By using smaller zenith angles (e.g. locally in a three-dimensional atmospheric model) one could still obtain larger ratios of $F_\mathrm{a,c}/F_\mathrm{a,cs}$ resulting also in a higher scattering greenhouse effect.

\begin{acknowledgements}
  The authors thank Gary Hansen for providing his refractive index of CO$_2$ ice and Tom Rother as well as Philip von Paris and Mareike Godolt for the fruitful discussions. Additionally, we also thank the referee Francois Forget for his suggestions improving the manuscript. This work has been partly supported by the research alliance \textit{Planetary Evolution and Life} of the Helmholtz Association (HGF).
\end{acknowledgements}

\bibliographystyle{aa} % style aa.bst
\bibliography{AA_2012_20025}

\end{document}